\documentclass[prl,twocolumn]{revtex4}
\usepackage{amsmath,latexsym,amssymb}
\usepackage[dvips]{graphicx}
\usepackage{epsfig}

\def\vec#1{\mathchoice
 {\mbox{\boldmath $\displaystyle#1$}}
 {\mbox{\boldmath $\textstyle#1$}}
 {\mbox{\boldmath $\scriptstyle#1$}}
 {\mbox{\boldmath $\scriptstyle#1$}}}

\begin{document}

\title{Self-consistent relativistic
random phase approximation \\with vacuum polarization}

\author{
A. Haga$^{1,2}$\footnote{Electronic address:
haga.akihiro@nitech.ac.jp}, H. Toki$^1$\footnote{Electronic
address: toki@rcnp.osaka-u.ac.jp}, S.
Tamenaga$^1$\footnote{Electronic address:
stame@rcnp.osaka-u.ac.jp}, Y. Horikawa$^3$\footnote{Electronic
address: horikawa@sakura.juntendo.ac.jp}, and H. L.
Yadav$^4$\footnote{Electronic address: hlyadav@sancharnet.in} }
\affiliation{ $^1$Research Center for Nuclear Physics (RCNP),
Osaka University, Ibaraki, Osaka 567-0047, Japan\\
$^2$Department
of Engineering Physics, Electronics, and Mechanics,
Nagoya Institute of Technology, Gokiso, Nagoya 466-8555, Japan\\
$^3$Department of Physics, Juntendo University, Inba-gun, Chiba
270-1695, Japan\\
$^4$Physics Department, Rajasthan University, Jaipur-302004, India}



\begin{abstract}

We present a theoretical formulation for the description of
nuclear excitations within the framework of relativistic
random-phase approximation whereby the vacuum polarization arising
from nucleon-antinucleon fields is duly accounted for. The vacuum
contribution to Lagrangian is explicitly described as extra new
terms of interacting mesons by means of the derivative expansion
of the effective action. It is shown that the
self-consistent calculation yields zero eigenvalue for the
spurious isoscalar-dipole state and also conserves  the
vector-current density.

PACS number(s): 21.10.-k,21.60.Jz, 13.75.Cs
\end{abstract}

\maketitle

\section{Introduction}

The relativistic field theory based on the quantum hadrodynamics
(QHD)\cite{WA74} has been very successful in describing the
nuclear properties not only for ground states but also for excited
states. Although the response of a system to an external field has
been already investigated in the eighties by using the
relativistic random-phase approximation (RRPA) in the relativistic
mean-field (RMF) basis
\cite{FU85,KUSU85,IC87,WE87,HOPI88,FU89,MC89}, the self-consistent
methods with nonlinear effective Lagrangian for a quantitative
description of excited states have been developed only during the
last few years \cite{MA97,CO98,PI00,RI01,GR03}. However, in particular,
it is important to emphasize here that the negative-energy RMF
states contribute essentially to the current conservation of RRPA
eigenstates and the decoupling of the spurious state. In our
recent study, it has been shown that the negative-energy RRPA
eigenstates generated from the RRPA equation with the fully
consistent basis have a significant role for gauge invariance in
the electromagnetic response\cite{HA041}.

Almost all investigations of RRPA in the QHD model referred to
above, however, neglected actual antinucleon degrees of freedom.
The basis set used for the RRPA calculation is usually obtained
from the RMF theory wherein only positive-energy nucleons are
taken into account and the Dirac sea is always regarded as
unoccupied (this is sometimes called {\it no-sea approximation}).
This approximation is very convenient because we do not have to
puzzle over a renormalization procedure not only in the
calculation of basis set under the full one-nucleon-loop
contribution, which we refer to as the relativistic Hartree
approximation (RHA), but also in the RRPA calculation in which the
Feynman part is essentially divergent. Thus, the RRPA calculations
in a finite nuclear system with the inclusion of vacuum
polarization have been performed only approximately
\cite{IC87,WE87,HOPI88,FU89}. All the RHA + RRPA calculations
performed earlier have been based on the local-density
approximation, that is, the renormalization of the one-nucleon
loop in the RHA calculation and of the Feynman part in the
polarization function were done in nuclear matter and the results
were applied to finite nuclei. 
However, as the local-density approximation of the
nucleon-antinucleon loop corrections violates the self-consistency
for finite nuclei, the spurious isoscalar-dipole strength
associated with the uniform translation of the center-of-mass dose
not get shifted all the way down to zero excitation
energy\cite{HOPI88}.

The main aim of this letter is to demonstrate as to how such a
deficiency contained in the previous RHA + RRPA calculations is
removed by employing the derivative-expansion method to estimate
the vacuum contribution. We verify the consistency by calculating
explicitly the spurious state in the isoscalar-dipole mode and the
current conservation of transition densities.

In the following we work out the vacuum-polarization effects using
the Lagrangian density of the Walecka $\sigma-\omega$ model which
is given by
\begin{eqnarray}
\label{nucleusL}
{\cal L}_N &=&
\bar{\psi}_N(i\gamma_\mu \partial^\mu-m_N+g_\sigma\sigma-g_\omega\gamma^\mu\omega_\mu)\psi_N
\nonumber\\
&&
+\frac{1}{2}(\partial_{\mu}\sigma)^2
-\frac{1}{2}m_{\sigma}^2\sigma^2
-U(\sigma)
\\
&&
-\frac{1}{4}(\partial_{\mu}\omega_{\nu}-\partial_{\nu}\omega_{\mu})^2
+\frac{1}{2}m_{\omega}^2\omega_\mu\omega^\mu
-\delta {\cal L},
\nonumber
\end{eqnarray}
where
$U(\sigma)=\frac{1}{3!}g_{2}\sigma^3+\frac{1}{4!}g_{3}\sigma^4$
denotes the self-interaction terms of scalar meson, and $\delta
{\cal L}=-\frac{1}{4}\zeta_\omega
(\partial_{\mu}\omega_{\nu}-\partial_{\nu}\omega_{\mu})^2
+\frac{1}{2}\zeta_\sigma(\partial_{\mu}\sigma)^2 +\sum_{i=1}^4
(\alpha_i/i!)\sigma^i$ represents the counterterms to regularize
the nucleon self-energy. The renormalization procedure in a finite
nuclear system requires considerable effort even at the mean-field
level\cite{HA042,ST97}. The explicit calculation of the vacuum
polarization for finite system has been performed recently by Haga
et. al.\cite{HA042} wherein the density variation has
been found to be substantially large. The use of the effective action
developed in Ref.~\cite{JA74}, on the other hand, provides
a direct and simple approach to estimate the
vacuum corrections. It is interesting to observe that the lowest order of
derivative expansion for the one-baryon-loop correction
agrees with the rigorous results\cite{HA042}.
In the derivative expansion, the
contribution from the Dirac sea to the Lagrangian density, which
is expressed by the trace and the logarithm of the inverse of the
Dirac Green function, is expanded by derivatives of the meson
fields as expressed by
%
%
\begin{eqnarray}
\int \frac{d^4p}{(2\pi)^4}[ Tr\ln(\gamma_\mu p^\mu&-&m_N+g_\sigma\sigma
-g_\omega\gamma^\mu\omega_\mu)
\nonumber
\\
&-&Tr\ln(\gamma_\mu p^\mu-m_N)]
-\delta {\cal L}
\nonumber
\\
=
-V_F(\sigma)+\frac{1}{2}Z_F^\sigma(\sigma)(&\partial_\mu& \sigma)^2
\nonumber
\\
+\frac{1}{4}Z_F^\omega(\sigma)
(&\partial_{\mu}&\omega_{\nu}-\partial_{\nu}\omega_{\mu})^2+\cdots.
\label{DE}
\end{eqnarray}
Each term of the right-hand side of Eq.(\ref{DE}) is finite, and
therefore we can treat them as ordinary potential terms.
Here, we may need to use a standard technique of the
renormalization to obtain explicit forms of $V_F(\sigma)$,
$Z_F^\sigma(\sigma)$, and $Z_F^\omega(\sigma)$. The method of
calculating them has been discussed by many
authors\cite{AI84,CH85,OCH85} and it has also been verified that
the convergence of the expansion is quite rapid within a mean-field
approximation\cite{HA042,ST97,PE86}. In the present calculation, we
employ only the first three terms in the right-hand side of
Eq.(\ref{DE}).

We proceed now to describe the calculational details of the
random-phase approximation based on the leading-order terms of the
derivative expansion. In the $\sigma-\omega$ model, we know that
the effective Lagrangian density is given by
\begin{eqnarray}
\mathcal{L}^{ren}_{(1)}&=&\bar{\psi}_N^+ (i\gamma_\mu
\partial^\mu-m_N+g_\sigma\sigma-g_\omega\gamma^\mu\omega_\mu)\psi_N^+
\nonumber\\
&&+\frac{1}{2}(\partial_\mu\sigma)^2-\frac{1}{2}m_\sigma^2\sigma^2
-U(\sigma)
\nonumber\\
&&-\frac{1}{4}(\partial_{\mu}\omega_{\nu}-\partial_{\nu}\omega_{\mu})^2
+\frac{1}{2}m_\omega^2(\omega_{\mu})^2
\\
&&-V_F(\sigma)+\frac{1}{2}Z_F^\sigma(\sigma)(\partial_\mu \sigma)^2
\nonumber\\
&&+\frac{1}{4}Z_F^\omega(\sigma)
(\partial_{\mu}\omega_{\nu}-\partial_{\nu}\omega_{\mu})^2,
\nonumber
\label{LDE}
\end{eqnarray}
where superscript $+$ in the nucleon-field operators means that
only the positive-energy states are to be treated explicitly.
Then, assuming the stationary and spherical system, the ground
state expectation values of the $\sigma$ and $\omega_0$, which are
written as $\phi$ and $V_0$, satisfy the following coupled equations:
\begin{align}
(\partial_\mu\partial^\mu+m_\sigma^2)\phi&=g_\sigma\langle \bar{\psi}^+\psi^+ \rangle
-U^\prime(\phi)-V_F^{\prime}(\phi)
\nonumber
\\
&+\frac{1}{2}Z_F^{\sigma \prime}(\phi)(\partial_\mu \phi)^2
-\partial_\mu (Z_F^\sigma(\phi) \partial^\mu \phi)
\nonumber\\
&+\frac{1}{4}{Z_F^{\omega}}^{\prime}(\phi)(\partial_{\mu}V_0)^2,
\\
(\partial_\mu\partial^\mu+m_\omega^2)V_0&=g_\omega\langle \bar{\psi}^+\gamma^0\psi^+ \rangle
\nonumber\\
&+\partial_\mu (Z_F^\omega(\phi) \partial^\mu V_0),
\end{align}
where the contributions of the negative-energy states to source terms
are contained in $V_F$ and $Z_F$.
The first step in calculating the RRPA response is the computation
of the RHA, which is just to solve these equations together with
the Dirac equation under the potentials of $\phi$ and $V_0$.
The potentials achieved in the mean-field calculation are
then used to generate the lowest-order polarization function.
We mention here that the $V_F$ and $Z_F$ terms, which appear in the meson propagator
parts, correspond to the excitation of negative-energy states to all the
positive-energy states.
Thus, we solve the RRPA equation given by
\begin{align}
\Pi_{RPA}(\Gamma^a&,\Gamma^b;\vec{p},\vec{q};E)=\Pi_D(\Gamma^a,\Gamma^b;\vec{p},\vec{q};E)
\\
+&\sum_i g_i^2
\int d\vec{k}_1 d\vec{k}_2
\Pi_D(\Gamma^a,\Gamma^i;\vec{p},\vec{k}_1;E)&
\nonumber
\\
&\times D_i(\vec{k}_1,\vec{k}_2;E) \Pi_{RPA}(\Gamma^i,\Gamma^b;\vec{k}_2,\vec{q};E),
\nonumber
\label{RPAeq}
\end{align}
where the summation is over the meson fields and $\Gamma$'s are
the $4 \times 4$ matrices which denote the vertex couplings.
This RRPA equation exactly has the same form as that of {\it no-sea
approximation}. However, it should be noted that there is an
essential difference in regard to the employed meson propagator,
$D_i$, in which we {\it must} include the vacuum-polarization
corrections to perform the consistent calculation.
The $\sigma-\omega$ coupling term
$Z_F^\omega(\sigma)(\partial_\mu\omega_\nu-\partial_\nu\omega_\mu)^2$
in the present Lagrangian provides the coupled equations between
the $\sigma$-meson propagator and the time component of the
$\omega$-meson propagator
\begin{align}
\left(
\begin{array}{cc}
D_{\sigma} & D_{\sigma\omega} \\
D_{\omega\sigma} & D_{\omega00} \\
\end{array}
\right)
=&
\left(
\begin{array}{cc}
D_{\sigma}^0 & 0\\
0 & D_{\omega00}^0 \\
\end{array}
\right)
\nonumber\\
+
\left(
\begin{array}{cc}
D_{\sigma}^0 & 0\\
0 & D_{\omega00}^0 \\
\end{array}
\right)&
\left(
\begin{array}{cc}
\tilde{\Pi}_F^{\sigma\sigma} & \tilde{\Pi}_F^{\sigma\omega} \\
\tilde{\Pi}_F^{\omega\sigma} & \tilde{\Pi}_F^{\omega\omega} \\
\end{array}
\right)
\left(
\begin{array}{cc}
D_{\sigma} & D_{\sigma\omega} \\
D_{\omega\sigma} & D_{\omega00} \\
\end{array}
\right),
\end{align}
while the spatial component of $\omega$-meson propagator can be evaluated
independently.
%
%
%
%
%
Here, the Fourier transforms of
\begin{align}
\tilde{\Pi}_F^{\sigma\sigma}(y,x)&= \delta(x-y)
\left[
U^{\prime\prime}(\phi)
+V_F^{\prime\prime}(\phi)
\right.
\nonumber\\
&-\frac{1}{2}Z_F^{\sigma \prime\prime}(\phi)(\partial_\mu \phi)^2
-\frac{1}{2}Z_F^{\omega \prime\prime}(\phi)(\partial_\mu V_0)^2
\nonumber\\
&
\left.
+ ( \partial_\mu \partial^\mu Z^\sigma_F(\varphi) ) 
\right]
+ (\partial^\mu)\left[Z_F^\sigma(\phi)\left[\partial_\mu \delta(x-y)
\right]\right]
\label{piss}
\\
\label{piws}
\tilde{\Pi}_F^{\sigma\omega}(y,x)& 
= \partial_\mu[Z_F^{\omega\prime}(\phi) (\partial_\mu V_0)\delta(x-y)]
\\
\label{pisw}
\tilde{\Pi}_F^{\omega\sigma}(y,x) &
= - Z_F^{\omega\prime}(\phi) (\partial_\mu V_0)[\partial^\mu \delta(x-y)] 
\\
\label{piww}
\tilde{\Pi}_F^{\omega\omega}(y,x)&= 
\partial^\mu \left[ Z_F^\sigma(\phi)\left[\partial_\mu \delta(x-y)
\right] \right]
\end{align}
and the effective meson propagators $D_i$ as well as free meson
propagators $D_i^0$ are expressed as the matrices of the momentum
space.
Details of the formulation on
the present technique along with the comparison between the
Feynman part in our calculation and that in the local-density
approximation will be elucidated in a forthcoming publication.

The parameter set used in the present work is listed in Table I
and has been determined to reproduce the total binding energies
and charge radii of the spherical nuclei in the RHA calculation.
This parameter set is similar to that introduced in
Ref.~\cite{MA99}, where the derivative expansion has been used.
Small difference between the two sets stems from the fact that in
the treatment of Ref.~\cite{MA99} the one-meson loop corrections
have been considered in the calculation of the ground state.
\begin{table}
\caption[tab1]
{\footnotesize
Parameter set used in the present work.}
\begin{math}
\begin{array}{cccccccccc}
\hline\hline
&\hspace{0mm} m_{\sigma}({\rm MeV}) \hspace{0mm}
&\hspace{0mm} m_{\omega}({\rm MeV}) \hspace{1mm}
&\hspace{1mm} g_{\sigma} \hspace{1mm}
&\hspace{1mm} g_{2} ({\rm fm}^{-1}) \hspace{0mm}
&\hspace{1mm} g_{3} \hspace{0mm}
&\hspace{1mm} g_{\omega} \hspace{0mm}\\
\hline
{\rm RHA} & 458.0 & 814.0 & 7.10  & 24.09 & -15.99 & 8.85\\
\hline\hline
\end{array}
\end{math}
\end{table}

In what follows we discuss the results of our RRPA calculations.
In Figs.~1(a) and 1(b) we display the distributions for the
Coulomb responses of isoscalar-dipole mode in $^{16}$O and
$^{40}$Ca at the momentum transfers of $q=237$ MeV and $q=118$
MeV, respectively as the function of the excitation energy. The
most remarkable result of our RRPA calculations including the
vacuum polarization is that the spurious state which is the
collective mode corresponding to the center-of-mass motion,
appears at zero excitation energy in both nuclei (shown by solid
curves) as also obtained in the conventional RRPA calculation
without vacuum polarization (shown by dashed curves)\cite{FU85}.
The spurious state is clearly separated from the physical states,
which are shown by using the small imaginary part of energy
$\eta=0.05$ MeV. We emphasize that although earlier RRPA
calculations with vacuum polarization have never succeeded to
decouple the spurious state\cite{HOPI88}, we are now able to
achieve this by handling the Lagrangian (\ref{LDE}) correctly.

\begin{figure}[h]
\centering
\includegraphics[width=7.0cm,clip]{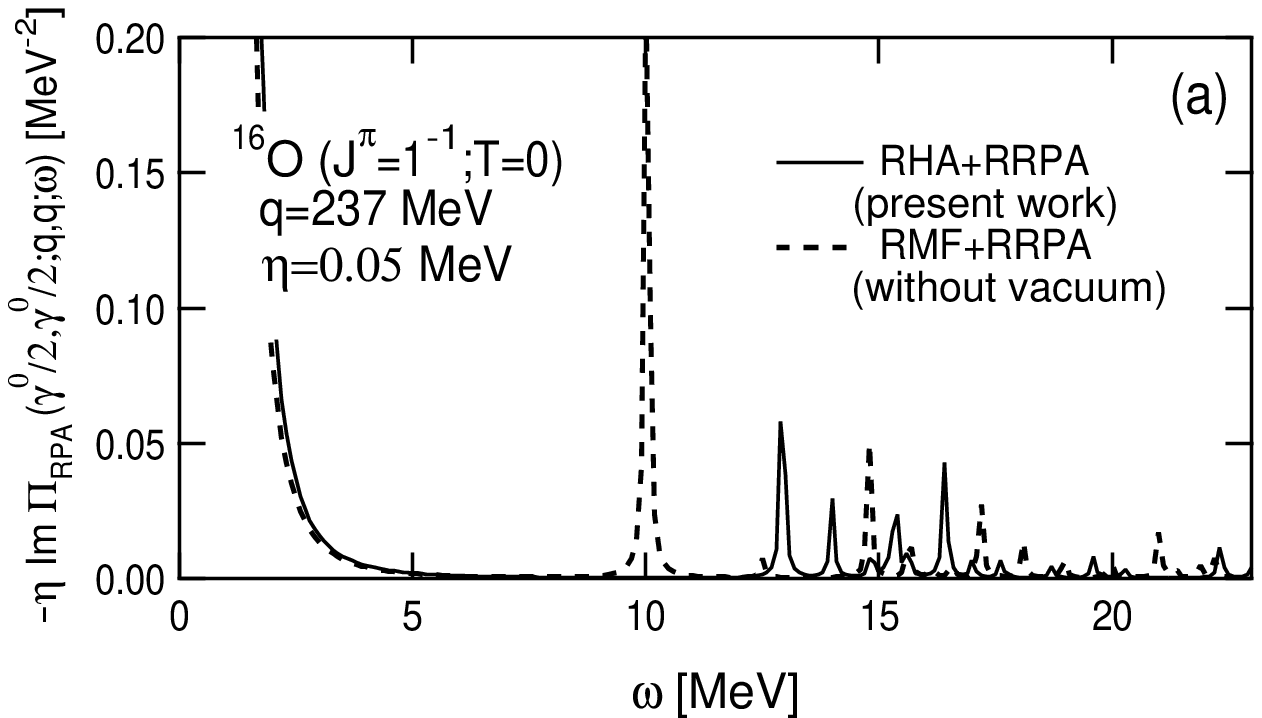}
\includegraphics[width=7.0cm,clip]{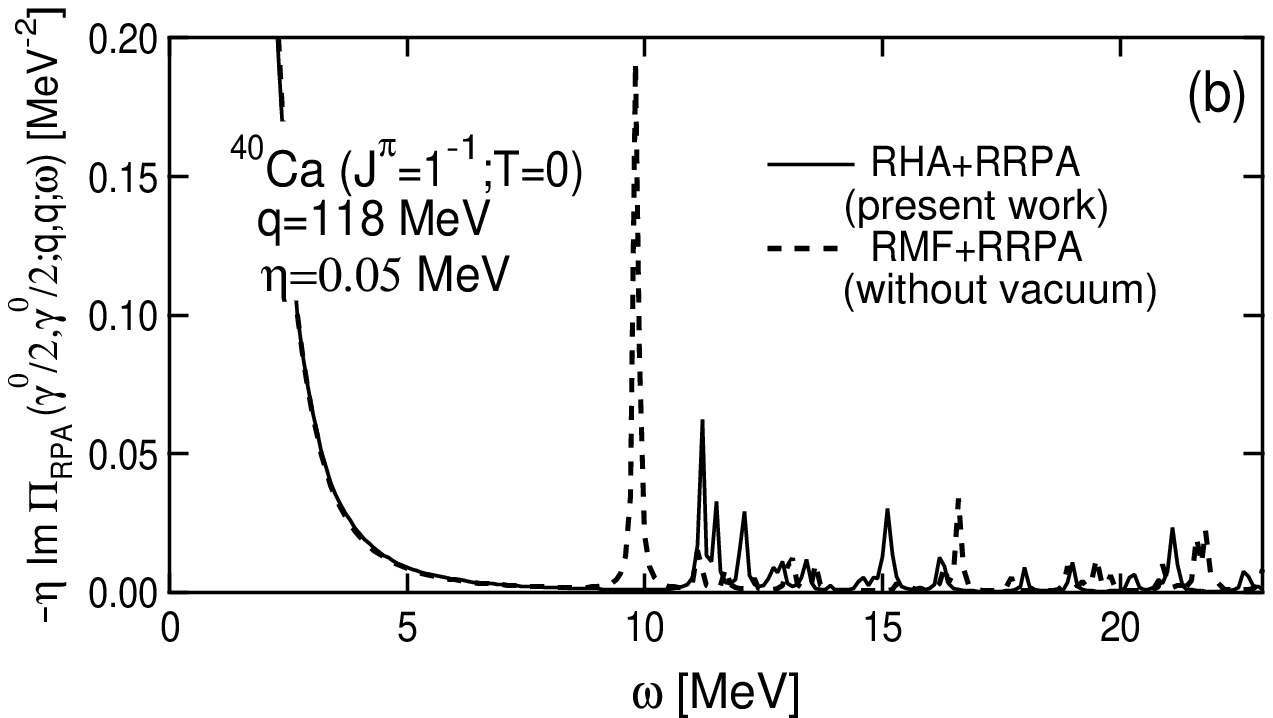}
\caption{\label{fig1}
Distributions of isoscalar-dipole strength in (a) $^{16}$O
and in (b) $^{40}$Ca.}
\end{figure}

\begin{figure}[h]
\centering
\includegraphics[width=7.0cm,clip]{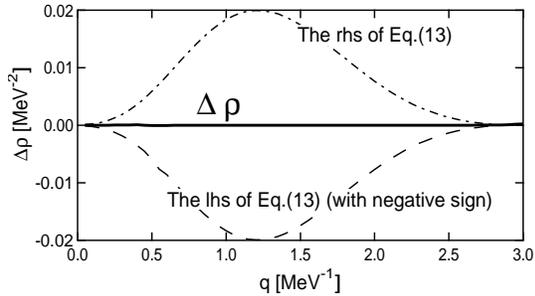}
\caption{\label{fig2} RHA + RRPA results for the current
conservation in the transition densities for the lowest
positive-energy (12.94MeV) isoscalar-dipole state in $^{16}$O. The
$\Delta\varrho$ depicted by thick-solid line shows the extent of
violation of current conservation. For details refer to the text.}
\end{figure}

Another important aspect for the correctness of the RPA
calculations is to check whether transition charge density
$\langle I'|| M_\lambda(q)||I \rangle$ and current density
$\langle I'|| T_{\lambda L}(q)||I \rangle$, connecting the ground
state $I$ and the excited states $I'$ for different multipolarity
$\lambda$ and $L$, satisfy the conservation law \cite{FU85,MC89}.
Assuming $\omega_N$ denotes the excitation energy of the nucleus,
the conservation relation is given by,
\begin{align}
\omega_N\langle I'|| M_\lambda(q)||I \rangle&=
-q \sqrt{\frac{\lambda}{2\lambda+1}}
\langle I'|| T_{\lambda\lambda-1}(q)||I \rangle
\nonumber
\\
+&q
\sqrt{\frac{\lambda+1}{2\lambda+1}}
\langle I'|| T_{\lambda\lambda+1}(q)||I \rangle.
\label{conservation}
\end{align}
Our results for the RRPA transition current have been depicted in
Fig.~2 where the contributions from the left-hand side (lhs) and
right-hand side (rhs) of Eq.~(\ref{conservation}) are shown
separately by the dashed and dash-dotted lines, respectively. Note that
for the purpose of clarity the lhs of Eq.~(\ref{conservation}) has
been plotted with a negative sign. The difference of these two
contributions denoted by $\Delta \varrho$ and shown by thick-solid
line in Fig.~2, represents the violation of the current
conservation. It is gratifying to see from Fig.~2 that the RRPA
transition current is sufficiently conserved.

In summary, we have studied the self-consistent RHA + RRPA method
including the vacuum-polarization contribution given by the
derivative-expansion method. We stress here that the present
method is able to treat the change of the negative-energy states
due to the presence of particles in the positive-energy states for
the formation of the nucleus and also to treat consistently
excited states using the RRPA. In contrast to the
previous calculations based on the local-density approximation,
the self-consistency of the model has been fulfilled so that we
have obtained the desired results for decoupling the spurious
isoscalar-dipole state, and for conserving the current density.

Also, it would be pertinent to take stock of the present scenario
as regards to where we stand now. Indeed we now have a powerful
method of describing the ground state of nuclei in the
relativistic mean-field approximation with the inclusion of
negative-energy states which are influenced by the meson mean
fields. Further, we also now have a method of calculating the
excited states in nuclei whereby the excitation of nucleons from
negative-energy states to positive-energy states is duly included.
This makes the present treatment of excited states fully
consistent with the description of the ground state. Numerically,
however, the $\sigma$- and $\omega$-mean fields obtained in the
present treatment are  found to be insufficient in strength,
entailing a weaker spin-orbit splitting which is about a half as
compared to that obtained in the usual RMF model \cite{SU94}.
There are several possibilities to provide the missing spin-orbit
splitting. One interesting idea is to explore  the possibility of
surface pion condensation suggested recently \cite{OG04}. Another
thing could be to consider the tensor coupling for the $\omega$
meson \cite{SU942,MA03}. These extensions of the model constitute
an effective field theory including the vacuum polarization. 
Further calculations testing the RRPA with these
extended RHA models for a wide variety of observables in nuclear
excitations would be of immense interest.

This work has been supported by MATSUO FOUNDATION, Suginami,
Tokyo. We thank Prof. P. Ring and Prof. A. E. L. Dieperink for
fruitful discussions.




\end{document}